\documentclass[conference]{IEEEtran}
\IEEEoverridecommandlockouts
\usepackage{amsmath,amssymb,amsthm,graphicx,booktabs,xcolor,cite,enumitem}

\newtheorem{remark}{Remark}

\newtheorem{proposition}{Proposition}
\setlist{nosep,leftmargin=*}
\begin{document}
\title{Job-Level Workload Allocation for Data Centers Under Flexible Interconnections\\
\thanks{This material is based upon work supported by the U.S. Department of Energy’s Office of EERE under award number DE-EE0010147. The views expressed herein do not necessarily represent the views of the U.S. Department of Energy or the United States Government.}
}
\author{\IEEEauthorblockN{ Tanmay Mishra, Arash Omidi, Mads R. Almassalkhi} \IEEEauthorblockA{\textit{Department of Electrical and Biomedical Engineering, University of Vermont} \\Burlington, Vermont, USA \\ \{tmishra, aomidi, malmassa\}@uvm.edu}  }

\maketitle

\begin{abstract}
Flexible interconnections allow data centers to connect to the grid sooner without upgrading the grid infrastructure, if they accept limits on their grid power during periods of grid stress. When a limit is reached, the data center must slow some of its computing jobs. This choice sets the cost of flexibility, because the work done by a job is worth far more than the energy it uses. This paper models how jobs respond when their power is reduced, and proposes a real-time method that shares the available power among jobs according to their value. It also gives grid planners a simple measure of how much power a data center can reduce, and for how long. A one-year case study of a 100-MW data center shows that the proposed method lowers the cost of meeting grid limits, and that on-site generation and storage lower it further.
\end{abstract}
\vspace{-0.25cm}
\section{Introduction}
Data centers have rapidly emerged as the defining large loads of the modern electric grid, driven by the exponential growth of cloud computing and artificial intelligence. Unlike conventional loads, a large data center facility can request hundreds of MW at one location, and the network upgrades needed to serve it can take years. Recent North American Electric Reliability Corporation (NERC) assessments identify gaps in modeling emerging large loads, evaluating their interconnection impacts, and coordinating their operation with grid operators~\cite{NERC_2026_LargeLoads}. Further, the Federal Energy Regulatory Commission (FERC) directed regional grid operators to justify or revise their large-load interconnection rules~\cite{FERC_RM26_4}. 

In this context, one alternative to waiting for network upgrades is flexible large-load service, typically referred to as flexible interconnection. It allows large loads to obtain earlier grid access by being \emph{Good Grid-citizen} in exchange for temporarily reducing electricity imports when needed~\cite{kahrl2026speed}. The potential is substantial as observed by a recent Duke University study, which estimates that the existing U.S. grid could accommodate up to 98 GW of additional flexible load with an average annual curtailment rate of just 0.5\%, involving relatively short and infrequent curtailment events~\cite{norris2025rethinking}.


This trade is attractive only if curtailment is cheap for the data center. A data center's computing work is worth far more than its electricity e.g., NVIDIA~H100~SXM rental rates of approximately \$3–5/GPU-hour correspond to \$4,300–7,100/MWh when normalized by rated power consumption (roughly 700~W per GPU), compared with approximately \$85/MWh for U.S. industrial electricity \cite{eia_electricity_price}. What matters is therefore not how many MWh are curtailed, but which work is lost. Training and batch jobs can be slowed, because their deadlines are hours away~\cite{radovanovic2023}, while interactive inference cannot be flexible. GPUs allow this through a power cap~\cite{you2023zeus}. However, jobs respond very differently to the same cap, e.g., some lose most of their speed, others very little~\cite{you2023zeus}. Jobs also differ in value and in how much delay their Service Level Agreement (SLA) allows. The same MW of curtailment can therefore cost very different amounts, depending on which jobs absorb it.

Current practice and models do not capture this. Data centers throttle by priority, cutting lower-priority work first~\cite{thunderbolt}, and a recent field test curtailed an AI cluster on a utility request by slowing and pausing jobs~\cite{colangelo2507turning}. Grid-side studies represent the data center as an aggregate flexible load, often a virtual battery~\cite{wang2019frequency}. Neither view tells the operator which jobs should give up power, nor tells the grid planner how long a site can sustain a limit. Data centers are also adding their own generation and storage, often called \emph{bring-your-own capacity} (BYOC). FERC has directed PJM to set rules for large loads co-located with generation~\cite{FERC_2025_PJM_CoLocation}. Hybrid plants that combine generation and storage at one connection point are growing quickly. Most data centers already install backup generators for reliability~\cite{brooks2025datacenter}. Under flexible interconnection, BYOC can cover a grid limit before any job slows down, which raises the very critical question of how to combine them with job curtailment.

\begin{figure*}[!htb]
\centering
\includegraphics[width=0.95\textwidth]{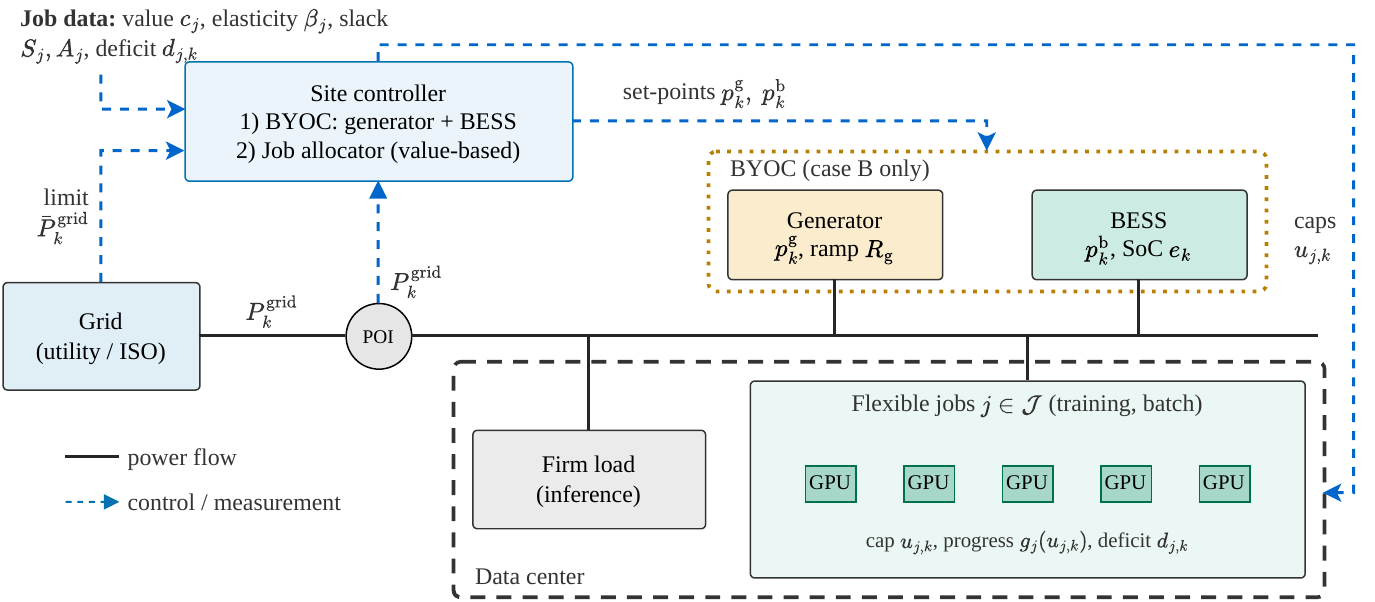}
\caption{Data center under a flexible interconnection where grid sends the import limit $\bar P^{\mathrm{grid}}_k$, the site controller dispatches the power limits to set up the GPU power cap $u_{j,k}$ of each flexible job $j$ and  power set point for on-site generation and storage (BYOC case).}
\label{fig:system}
\vspace{-0.3cm}
\end{figure*}

This paper addresses these questions. Its contributions are:
\begin{enumerate}
\item A job-level model of a data center under a grid import limit. Each job has a power cap, a speed sensitivity, a value, and an SLA slack. 
\item We proposed a real-time, value-based job allocation that is unique and optimal at each step, benchmarked against uniform and priority-based job allocation.
\item A closed-form shedding capability curve that gives the power, a site can shed w.r.t event duration, and the recovery time.
\end{enumerate}
The remainder of the paper is organized as: Section~\ref{sec:model} presents the system model, Section~\ref{sec:alloc} the proposed allocation policy, Section~\ref{sec:case} the case study, and Section~\ref{sec:conclusion} concludes the paper.

\section{System Model}\label{sec:model}
In this work, the system is a data center connected to the grid as shown in Fig.~\ref{fig:system}. At the point of interconnection (POI), the grid power ($P^{\mathrm{grid}}_k$) is measured. Behind the POI, the site serves two kinds of load: a) firm load, $P^{\mathrm{firm}}_k$ and b) flexible load, $P^{\mathrm{flex}}_k$. The firm load runs interactive services such as AI inference and it can't be curtailed. The flexible load runs training and batch jobs on GPUs, each under its own power cap. In case B, the site also has an on-site generator with output $p^{\mathrm g}_k$ and a BESS with dispatch $p^{\mathrm b}_k$ (negative when charging) under the BYOC consideration. The flexible interconnection requires
\begin{equation}
P^{\mathrm{grid}}_k=P^{\mathrm{firm}}_k+P^{\mathrm{flex}}_k-p^{\mathrm g}_k-p^{\mathrm b}_k\le\bar P^{\mathrm{grid}}_k ,
\label{eq:limit}
\end{equation}
The grid operator sends the site an import limit $\bar P^{\mathrm{grid}}_k$. A site controller receives this limit, the measured grid power, and the state of each job. It acts in two stages. It first dispatches the generator and the BESS, if present. It then sets the GPU power cap $u_{j,k}$ of each flexible job so that the grid power stays below the limit for every time step $k$ of $\Delta_h$ hours. These GPU caps take effect within seconds, so the controller can follow the limit at every step. Without on-site resources (case A), $p^{\mathrm g}_k=p^{\mathrm b}_k=0$, and only the flexible jobs can absorb a limit.


\subsection{Flexible job model of the data center}
The \emph{flexible} part consists of jobs whose deadlines are hours or days away, so their progress can be shifted in time without affecting their users~\cite{radovanovic2023}. We denote this set of jobs by $\mathcal J=\{1,\hdots, N\}$. 
Each flexible job $j \in \mathcal{J}$ runs on a dedicated set of GPUs with total maximum power $P_j$. All GPUs of a job share one power cap $u_{j,k}\in[u_{\min},1]$, a fraction of $P_j$, because a distributed job runs in lockstep and its slowest GPU sets its speed. Its scheduled cap is $u^0<1$, so the job draws $P_ju^0$ as per the schedule allocation with total flex power $P^{\mathrm{flex}}_k=\textstyle\sum_jP_ju^0$. A job may also represent a group of jobs with similar attributes. Lowering the cap slows the job, while raising it speeds the job up. We model the job's speed relative to its schedule as
\begin{equation}
g_j(u_{j,k})=\big(u_{j,k}/u^0\big)^{\beta_j},\qquad \beta_j\in(0,1).
\label{eq:g}
\end{equation}
The parameter $\beta_j$ is the job's speed sensitivity to power e.g., halving the cap of a job with $\beta_j=0.3$ leaves it at 81\% of its speed. It plays the role of an efficiency curve. A job that mostly waits for memory or data has a small $\beta_j$ i.e. barely slows down. However, a job limited by computation has $\beta_j$ near 1. Since $\beta_j<1$, speed falls less than power, as measured for GPU power caps~\cite{you2023zeus}. The form~\eqref{eq:g} is our modeling assumption where $\beta_j$ can be fitted to measured throughput at a few caps.

Let $d_{j,k}$ (h) be the schedule deficit of job $j$, the work it has fallen behind when slowed, with negative values meaning it is ahead. It evolves as
\begin{equation}
d_{j,k+1}=d_{j,k}+\Delta_h\big(1-g_j(u_{j,k})\big),\quad -A_j\le d_{j,k}\le S_j,
\label{eq:dc}
\end{equation}
The SLA slack $S_j$ is the maximum delay the SLA allows, and maximum lead $A_j$ is the limited by the work queued for the job. 
\begin{remark}
Each job thus acts as a small virtual battery, with deficit $d_{j,k}$ as its state of charge and $S_j+A_j$ as its energy capacity. With $\beta_j=1$, unlimited caps, and pooled slack, the jobs reduce to the single virtual battery common in load-flexibility studies~\cite{Sun2023}. In general, they form a pack of unbalanced cells, limited by the first cell to reach its bound, and the job allocation acts as the battery management system. 
\end{remark}
Each job has a value $c_j$ (\$/MWh of scheduled GPU energy). A job beyond $S_j$ incurs an SLA penalty equal to its value per hour of delay~\cite{GARG2014}. 


\subsection{On-site generation  and storage} 
As shown in Fig.~\ref{fig:system}, the site also has a generator and a BESS in case B (BYOC). They follow the standard models given as:
\subsubsection{BESS model}
Let $e_k$ (MWh) be the stored energy and $p^{\mathrm{b}}_k$ (MW) the BESS power, positive on discharge. With $\Delta_h$ in hour, $[x]_+=\max(0,x)$ and $[x]_-=\max(-x,0)$, the BESS energy dynamics and operating limits are
\begin{equation}
e_{k+1}=e_k-\Delta_h\Big(\frac{[p^{\mathrm{b}}_k]_+}{\eta}-\eta\,[p^{\mathrm{b}}_k]_-\Big),\quad e_{\min}\le e_k\le e_{\max},
\label{eq:bess_soc}
\end{equation}
where $\eta$ the charging and discharging efficiencies, $\overline{P}_{\mathrm{b}}$ is the rated power, and $[e_{\min},e_{\max}]$ is the usable energy range of the BESS. The energy bound also restricts the instantaneous power range at step $k$ to
\begin{equation}
-\min\!\Big\{\overline{P}_{\mathrm{b}},\tfrac{e_{\max}-e_k}{\eta\Delta_h}\Big\}
\le p^{\mathrm{b}}_k \le
\min\!\Big\{\overline{P}_{\mathrm{b}},\tfrac{(e_k-e_{\min})\,\eta}{\Delta_h}\Big\},
\label{eq:bess_range}
\end{equation}

\subsubsection{Generator model} Let $p^{\mathrm{g}}_k$ (MW) be the generator output with ramp rate $R_{\mathrm g}$ (MW/h) at time step $k$ and rated capacity of  $\bar P_{\mathrm g}$ (MW). The limits on the generator power is given by,
\begin{gather}
0\le p^{\mathrm g}_k\le\bar P_{\mathrm g},\qquad |p^{\mathrm g}_k-p^{\mathrm g}_{k-1}|\le R_{\mathrm g}\Delta_h,\label{eq:gen}
\end{gather}
Now, the total grid power can be rewritten as, $P^{\mathrm{grid}}_k=P^{\mathrm{firm}}_k+\sum_jP_ju^0-p^{\mathrm g}_k-p^{\mathrm b}_k-p^{\mathrm d}_k$, where $p^{\mathrm d}_k=\sum_jP_j(u^0-u_{j,k})$ representing the power the jobs give up with negative $p^{\mathrm d}_k$ means catch-up.

\section{Proposed Job Allocation Policy}\label{sec:alloc}
The site controller at every time step $k$, first computes the shortfall, the amount by which the scheduled draw exceeds the limit~\eqref{eq:limit} as shown in Fig.~\ref{fig:system}. In case A, the jobs absorb all of it. In case B, the generator covers it first within its ramp~\eqref{eq:gen}, and then the BESS within its power and energy~\eqref{eq:bess_soc}--\eqref{eq:bess_range}. The jobs absorb only the remainder,
\begin{equation}
\Delta_k=\Big[P^{\mathrm{firm}}_k+\textstyle\sum_jP_ju^0-p^{\mathrm g}_k-p^{\mathrm b}_k-\bar P^{\mathrm{grid}}_k\Big]_+ ,
\label{eq:delta}
\end{equation}
so job value is lost only when BYOC is exhausted. When the limit relaxes, the BESS recharges, the generator ramps down, and delayed jobs catch up. The policies below differ only in how they split $\Delta_k$ across jobs.

\subsection{Value-Based Real-Time Policy}
When the jobs absorb a shortfall $\Delta_k$, the power left for them is $B_k=\sum_jP_ju^0-\Delta_k$. The proposed policy shares $B_k$ using the \emph{value of a job's power},
\begin{equation}
v_{j,k}=c_j\,\beta_j\,w_{j,k},\qquad w_{j,k}=\frac{1}{\epsilon+[S_j-d_{j,k}]_+/S_j}.
\label{eq:vj}
\end{equation}
It combines the job's  $\beta_j$, its economic value $c_j$, and its urgency $w_{j,k}$. Urgency grows as the job uses its slack, and $\epsilon>0$ keeps it finite. A high $v_{j,k}$ marks a job that is valuable i.e. either slows down by a lot, or is close to its SLA limit.

At each step $k$, the controller chooses the caps $\mathbf u_k=(u_{j,k})_{j\in\mathcal J}$ of all flexible jobs. Its goal is the value of the work done, not the energy used. Each job contributes its value $c_j$ times its progress $g_j(u_{j,k})$, weighted by its urgency $w_{j,k}$, so that jobs close to their SLA limit are costly to cut. The grid limit leaves the jobs $B_k=\sum_jP_ju^0-\Delta_k$ MW, which they use in full. During a limit, each cap stays between the schedule $u^0$ and a lower bound. From~\eqref{eq:g}--\eqref{eq:dc}, the lowest cap that keeps job $j$ within its slack after the step is
\begin{equation}
\underline u_{j,k}=\max\Big\{u_{\min},\,u^0\big[1-(S_j-d_{j,k})/\Delta_h\big]_+^{1/\beta_j}\Big\}.
\label{eq:ulow}
\end{equation}
Now, allocation problem is formulated as,
\begin{equation}
\begin{aligned}
\max_{\mathbf u_k}\ &\textstyle\sum_j c_j\,w_{j,k}\,P_ju^0\,g_j(u_{j,k})\\
\text{s.t.}\ &\textstyle\sum_jP_ju_{j,k}=B_k,\quad \underline u_{j,k}\le u_{j,k}\le u^0 .
\end{aligned}
\label{eq:value}
\end{equation}
It is solved at every step, because grid limits are not known in advance.

\begin{proposition}\label{prop:opt}
For $\beta_j\in(0,1)$ and $u_{\min}>0$, problem~\eqref{eq:value} is convex. If $\sum_jP_j\underline u_{j,k}\le B_k$, its optimal job allocation is
\begin{equation}
u_{j,k}=\Pi_{[\underline u_{j,k},\,u^0]}\Big(u^0\big(v_{j,k}/\lambda_k\big)^{1/(1-\beta_j)}\Big),
\label{eq:rule}
\end{equation}
where $\lambda_k$ is the unique value at which $\sum_jP_ju_{j,k}=B_k$.
\end{proposition}
\begin{proof}
Each $g_j$ is strictly concave in its own cap, and the weights $c_jw_{j,k}$ are positive, so the objective is strictly concave. The constraints are affine, and the feasible set is closed, bounded and nonempty, so a unique solution exists. At this solution, every job strictly inside its bounds has the same marginal value of power. Per MW, job $j$'s marginal value is $v_{j,k}(u_{j,k}/u^0)^{\beta_j-1}$, with $v_{j,k}$ from~\eqref{eq:vj}. Setting it equal to $\lambda_k$, solving for $u_{j,k}$, and clipping the jobs at a bound gives~\eqref{eq:rule}. Each cap in~\eqref{eq:rule} decreases as $\lambda_k$ grows, so the total power is monotone in $\lambda_k$ and $\lambda_k$ is unique.
\end{proof}

Rule~\eqref{eq:rule} is our real-time job allocation policy. It satisfies the KKT conditions of~\eqref{eq:value}, and by Proposition~\ref{prop:opt} it is the unique job allocation. The site controller broadcasts one signal, $\lambda_k$, the marginal value of grid power, and each job sets its own cap. A job with a high value keeps a high cap, and a job that slows down little (low $\beta_j$) gives up more power. 

\begin{remark}
If the jobs cannot reach $B_k$ within their slack, every lower bound is relaxed to $u_{\min}$, and the SLA becomes soft for that step. If they cannot reach it even at $u_{\min}$, all jobs run at $u_{\min}$, and the remainder is an unavoidable limit exceedance. 
\end{remark}

Further, When the limit exceeds the scheduled draw, the spare power first recharges the BESS. It then lets delayed jobs ($d_{j,k}>0$) run faster than scheduled, up to $u=1$, by the same policy~\eqref{eq:rule} with caps in $[u^0,\bar u_{j,k}]$, where $\bar u_{j,k}\le1$ keeps each job within its maximum lead. This catch-up is the rebound that follows each event.



\subsection{Benchmark Policies}
We compare the proposed policy with two benchmarks. Each splits the same $\Delta_k$ across the same jobs. The \emph{Uniform capping} lowers every job by the same fraction $\theta_k$ of its range,
\begin{equation}
u_{j,k}=u^0-\theta_k\big(u^0-\underline u_{j,k}\big),
\label{eq:uniform}
\end{equation}
where $\theta_k\in[0,1]$ is chosen so that the jobs give up $\Delta_k$. It is the simplest rule and ignores all job differences.

\emph{Priority-based curtailment} follows current practice~\cite{thunderbolt,wu2016dynamo}. Jobs are ranked by value tier, $\pi_j=c_j$, from spot to standard to premium. The lowest tier is cut to its minimum before the next tier is touched. With $m_j$ the power taken from job $j$ and $\bar m_j$ its maximum, this is the exact solution of
\begin{equation}
\min_{0\le m_j\le\bar m_j}\ \sum_j\pi_jm_j\quad\text{s.t.}\quad\sum_jm_j=\Delta_k .
\label{eq:lp}
\end{equation}
Each MW taken from job $j$ costs a constant $\pi_j$, however far the job is already cut and however close it is to its SLA limit. In the value-based policy, this cost is $v_{j,k}(u_{j,k}/u^0)^{\beta_j-1}$, which rises as the job is cut and as its slack runs out. Priority therefore drains the cheapest tier to its floor, while the value-based policy spreads the cut and spares jobs near their SLA limit.

\section{Case Study}\label{sec:case}
For this study, we have used NLR's whole-facility data-center profiles at 1-min resolution over one year~\cite{nrel_dataset}, scaled to a 100-MW peak with means of 52~MW firm and 14~MW flexible. The flexible load is split into 16 job groups per day, with the parameters given in Table~\ref{tab:par}. This case study uses 1-min steps, $\Delta_h=1/60$~h. The value tiers mirror cloud pricing classes. The slack values bracket the event durations, and $u_{\min}=0.4$ is a conservative value within GPU power-limit ranges. The grid limit is 100~MW except during two kinds of events. Planned limits apply on summer weekdays from 16:00 to 20:00, 5~MW below the expected draw. About two contingencies per week last 30-120~min, 3-9~MW below the current draw. The annual cost is the lost work plus SLA penalties plus BYOC operation.


\begin{table}[!t]
\caption{Job and BYOC parameters}
\label{tab:par}
\centering\footnotesize
\begin{tabular}{@{}ll@{}}
\toprule
Speed sensitivity $\beta_j$ & $\mathcal U[0.3,0.8]$ \\
Scheduled / minimum cap & $u^0=0.8$, $u_{\min}=0.4$ \\
SLA slack $S_j$ & 0.25, 1 or 4~h; lead $A_j=0.5S_j$ \\
Value $c_j$ (spot, standard, premium) & \$1500, 3000, 6000/MWh \\
Generator (case B) & 5~MW,\, 36~MW/h, \$100/MWh \\
BESS (case B) & 2.5~MW,\, 5~MWh, $\eta=0.93$, \$30/MWh \\
\bottomrule
\end{tabular}
\end{table}

\subsection{Shedding Capability}
A grid planner must quantify the power a site can shed and how long it can sustain it, before agreeing to a flexible interconnection. A job held at cap $u$ falls behind at rate $1-g_j(u)$, from~\eqref{eq:dc}. Starting from deficit $d^0_j$, it stays within its slack for $T$ hours if $T(1-g_j(u))\le S_j-d^0_j$. The lowest cap that meets this condition gives the power the job can shed for $T$ hours, below its scheduled draw,
\begin{equation}
x_j(T)=P_j\Big(u^0-\max\Big\{u_{\min},\,u^0\big[1-\tfrac{S_j-d^0_j}{T}\big]_+^{1/\beta_j}\Big\}\Big),
\label{eq:env}
\end{equation}
and the site can shed $\sum_jx_j(T)$. Likewise, a job crosses its whole slack band, from $-A_j$ to $S_j$, in $T^{\mathrm c}_j=(S_j+A_j)/(1-g_j(u_{\min}))$ hours at $u_{\min}$, and recovers it in $T^{\mathrm r}_j=(S_j+A_j)/(g_j(1)-1)$ hours at $u=1$.

\begin{figure}[t]
\centering
\includegraphics[width=0.8\linewidth]{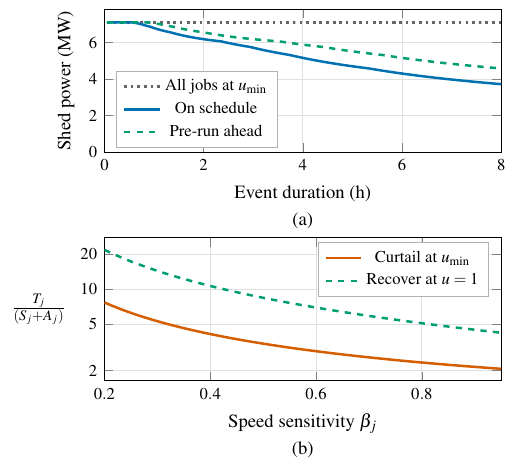}
\vspace{-0.5cm}
\caption{Shedding capability: (a) Power the site can shed versus event duration without SLA violations. (b) Curtailment and recovery times per hour of slack band, $T^{\mathrm c}_j/(S_j+A_j)$ and $T^{\mathrm r}_j/(S_j+A_j)$.}
\label{fig:env}
\end{figure}

Fig.~\ref{fig:env}(a) shows the curve for the average flexible load of 14.2~MW, with all jobs on schedule. The site can shed half its flexible load 7.1~MW, for about 30~min. For longer events, jobs with little slack must stay closer to $u^0$, so the curve falls to 6.8~MW for one hour and 5.2~MW for four hours. Running jobs ahead before a planned event ($d^0_j<0$) raises the curve, because each job then has more slack to spend. Fig.~\ref{fig:env}(b) shows that recovery is slower than curtailment. One hour of slack band takes 2.4--5.3~h to use at $u_{\min}$, but 5.1-14.4~h to recover at $u=1$, as the cap can drop 50\% below schedule but rise only 25\% above it in this case. A long event therefore leaves the site with less flexibility for the next one, so events need spacing unless $u^0= \frac{1+u_{\min}}{2}$.

\subsection{Policy Comparison}
\begin{figure}[!htb]
\centering
\includegraphics[width=0.8\linewidth]{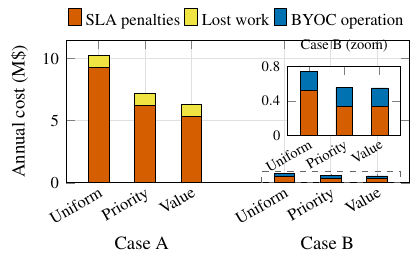}
\vspace{-0.5cm}
\caption{Annual cost by allocation policy.}
\label{fig:costs}
\vspace{-0.4cm}
\end{figure}

Fig.~\ref{fig:costs} and Table~\ref{tab:res} show comparison of three allocation policies for both the cases over one year data. In case A, every policy sheds the same 2380~MWh, so the policies differ only in which jobs pay for it. Value-based allocation lowers the annual cost by 12\% relative to priority and by 39\% relative to uniform capping, mainly through fewer SLA penalties. It also protects premium jobs slightly better than priority. Priority keeps cutting the cheapest tier until those jobs breach their SLA. The value-based policy also takes power from memory-bound or slack-rich jobs in higher tiers, which lose little.

BYOC matters more than the choice of policy. The generator and BESS cut the shed energy from 2380 to 218~MWh and the annual cost by about 90\%. Some limits cannot be met at all, because the firm load and the flexible floor exceed them. With BYOC, the policies differ less, since the jobs absorb little, but value-based allocation still has the lowest cost.

\begin{table}[!t]
\caption{Annual results, 100-MW data center}
\label{tab:res}
\centering\footnotesize
\setlength{\tabcolsep}{2.6pt}
\begin{tabular}{@{}lccc@{}}
\toprule
 & Uniform & Priority  & Value-based \\
\midrule
\multicolumn{4}{@{}l}{\emph{Case A: data center alone}}\\
SLA job-hours beyond slack & 2385 & 2093 & \textbf{1744} \\
Job-days with violation (\%) & 13.3 & 11.5  & \textbf{9.8} \\
Annual cost (M\$) & 10.28 & 7.20 & \textbf{6.32} \\
\midrule
\multicolumn{4}{@{}l}{\emph{Case B: data center + BYOC}}\\
SLA job-hours beyond slack & 147 & 130 & \textbf{124} \\
Annual cost (M\$) & 0.74 & 0.56 & \textbf{0.55} \\
\bottomrule
\end{tabular}
\vspace{-0.2cm}
\end{table}

\subsection{Event-Day Operation}
\begin{figure}[!htb]
\centering
\includegraphics[width=0.93\linewidth]{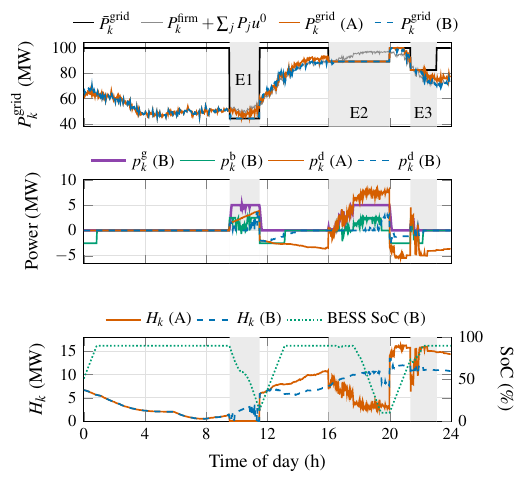}
\vspace{-0.25cm}
\caption{One summer weekday with three grid limits under case A and case B. Top, grid limit and grid import. Middle, generator, BESS and job power ($p^{\mathrm d}_k<0$ is catch-up). Bottom, job headroom $H_k$ and BESS state of charge.}
\label{fig:event}
\vspace{-0.3cm}
\end{figure}

Fig.~\ref{fig:event} shows one summer weekday with three limits. The headroom $H_k=\sum_jP_j(u_{j,k}-u_{\min})$ is the power the jobs could still shed at step $k$. The unplanned morning contingency E1 cuts the limit up to 13~MW below the scheduled draw. It arrives when the flexible load is at its daily low, so the data center alone exceeds the limit for two hours even with every job at $u_{\min}$. With BYOC, the generator and BESS supply up to 7.5~MW of the 13-MW shortfall, and the jobs cover most of the rest. During the planned limit E2, the data center alone can meet the limit, because the flexible load is high in the late afternoon, but its jobs absorb the full 20~MWh shortfall. With BYOC, the generator carries the sustained shortfall and the BESS covers its ramp and peaks, so the jobs absorb only 8\%. The evening contingency E3 binds for only 13 of its 103~min, because the evening load falls below the limit. Its 0.5-MWh shortfall is absorbed by the jobs in case A and almost entirely by BYOC in case B. Whether a limit is costly therefore depends on the site's load at that time.

Curtailed work returns after each event as rebound load. Without BYOC, delayed jobs draw 28.7~MWh above schedule later in the day, slightly more than the 26.0~MWh curtailed, because $\beta_j<1$. Curtailment therefore shifts energy rather than removing it. With BYOC, the rebound falls to 5.2~MWh. The BESS is nearly emptied in both E1 and E2, so its energy rating limits closely spaced events. In this case study, the job attributes are assumed, the NLR profiles are model-generated~\cite{nrel_dataset}, the firm load is treated as a known baseline, and the allocation does not plan for announced events.

\section{Conclusion}\label{sec:conclusion}
Flexible interconnection asks a data center to stay below a grid limit, and the cost of doing so depends on which jobs give up power. We modeled the data center at the level of its jobs and proposed a value-based allocation that equalizes the marginal value of each job's power and spares jobs near their SLA limit. The proposed allocation lowered the annual curtailment cost by 12\% relative to priority tiers in a one-year case study. By adding on-site generation and storage the cost is further reduced by 90\%. The shedding capability curve gives grid planners a simple measure of what a site can offer. Future work will extend this framework for data centers to provide various grid services such as frequency regulation. Further, the shedding capability curve can enter interconnection studies, to size flexible limits and BYOC for a given site.

\bibliographystyle{IEEEtran}
\bibliography{PESGM_Ref}
\end{document}